\documentclass[11pt,a4paper]{article}
\usepackage{jheppub_kim}

\usepackage{subfigure,amsmath,amssymb ,amsfonts, latexsym}

\usepackage[notcite,color,final]{showkeys}
\definecolor{refkey}{gray}{.25}
\definecolor{labelkey}{gray}{.25}

\def\be{\begin{equation}}
\def\ee{\end{equation}}
\def\bea{\begin{eqnarray}}
\def\eea{\end{eqnarray}}
\def\one{\mbox{1\kern-.59em{\rm 1}}}

 \vspace*{0cm}

\title{Vanishing Cosmological Constant by Gravitino-Dressed
Compactification of 11D Supergravity }

\author[a]{Fotis Farakos}
\author[a]{Alex Kehagias}
\author[a,b]{Emmanuel N. Saridakis}

\affiliation[a]{Physics Division, National Technical University of Athens,
15780 Zografou Campus,  Athens, Greece}

\affiliation[b]{CASPER, Physics Department, Baylor University, Waco, TX 
76798-7310, USA}

\emailAdd{fotisf@mail.ntua.gr}
\emailAdd{kehagias@central.ntua.gr}
\emailAdd{Emmanuel$_-$Saridakis@baylor.edu}


\abstract{We consider compactifications induced by the gravitino field  of eleven dimensional supergravity. 
Such compactifications are not trivial in the sense that the gravitino profiles are not related to pure bosonic ones 
by means of a supersymmetry transformation. The basic property of such backgrounds is that they admit  $\psi$-torsion 
although they have vanishing Riemann tensor. Thus, these backgrounds may be considered
also as solutions of the teleparallel formulation of supergravity. 
We construct two classes of solutions, one with both antisymmetric three-form field, gravity and gravitino and one with only gravity 
and gravitino. In these classes of solutions, the internal space is a
parallelized compact manifold, so that it does not inherit any 
cosmological constant to the external spacetime. The latter turns out to be flat Minkowski in the maximally symmetric case. 
The elimination of the cosmological constant in the spontaneously
compactified supergravity seems to be a generic property based on the trading of the cosmological constant for 
parallelizing torsion.  }

\begin{document}

\maketitle

\section{Introduction}

One of the most appealing ideas of modern theoretical physics is the
possibility of existence of extra  dimensions. Starting with the work of Kaluza and Klein, the idea that we live in a higher-dimensional
environment has been incarnated in string and M-theory as well as in the
brane-world scenarios. If this is the case and indeed
there are extra dimensions, large or small, or if we live in a 4D
hypersurface,  remains to be seen. 
A necessary part of every higher-dimensional  theory is how the theory compactifies, i.e., a four-dimensional
world is attained. Technically, compactification proceeds through the 
classical solutions, that  serve as a
vacuum of the field equations  and then by dimensional reduction  the
lower-dimensional effective theory is obtained \cite{Duff:1986hr}. 

In particular, the $N=1$ eleven-dimensional supergravity
\cite{Cremmer:1978km} is a maximal theory that offers a rich foundation
for studying 
compactifications and dimensional reduction to lower dimensions. 
There are many known compactifications of eleven-dimensional simple supergravity. For example, the Freud-Rubin \cite{Freund:1980xh}  compactification 
$AdS_4\times S^7$ or $AdS_7\times S^4$ gives rise to lower dimensional
theories on AdS spaces, much 
studied due to AdS/CFT correspondence \cite{malda}.
These AdS spaces are for instance holographic duals of gauge theories
but cannot be considered as true four-dimensional
vacua, mainly because they have a large negative cosmological constant. On the other hand, flat Minkowski four- or higher-dimensional vacua also
exist in eleven-dimensional supergravity when the latter compactifies on Ricci-flat internal spaces. 
We may recall the 4D ${\cal{N}}=8$ compactification on $T^7$ or the ${\cal{N}}=1$ 
compactification on $G_2$-holonomy manifolds, or even the ${\cal{N}}= 2$ compactification on $CY_3\times S^1$. Thus, we see that 
there are supersymmetric compactifications with flat 4D Minkowski
spacetime. One may ask of course if there are also non-supersymmetric 
compactifications with 4D Minkowski spacetime. For instance,
compactification on a generic non-Ricci flat internal space, 
e.g $Q(p,q,r)$  \cite{MQ}, will give rise to
an $AdS_4$ spacetime. The reason is that if the internal space is not Ricci-flat, it will have a positive cosmological constant (for 
a maximally symmetric 4D spacetime) which will give rise to a negative cosmological constant in 4D and therefor to $AdS_4$. 
Definitely, it is possible to consider an internal space with a negative
cosmological constant and finite volume (for example by
quotienting the internal space by a discrete appropriate subgroup of its
non-compact isometry group \cite{KR1,KR2,KR3,KR4}). In this case,
a de Sitter 4D vacuum emerges but with  a huge cosmological constant. 

Another possibility is to break supersymmetry by employing  different
boundary conditions for bosons and fermions. In a torus compactification
for example, one may  adopt antiperiodic boundary conditions for fermions
and periodic for bosons along the circles of the torus. This will give a
mass to the 4D gravitino of order the torus radii, which breaks 
supersymmetry and creates a cosmological constant proportional again to the
size of the internal space. Thus, we see that although there are
supersymmetric compactifications with flat 4D Minkowski vacuum,
non-supersymmetric ones are not-known or rare in the best case. It should
be stressed nevertheless, that in supersymmetric compactifications
Minkowski spacetime is no longer a real vacuum as soon as supersymmetry is
broken. Supersymmetry breaking produces a cosmological constant which
shifts the vacumm from Minkowski to de Sitter. We may then state that
compactifications of higher-dimensional supergravity, where only bosonic
fields are turned on, cannot result in a 4D Minkowski vacuum as long as
supersymmetry is broken. This problem is our motivation for searching for
non-supersymmetric flat 4D vacua in 11D simple supergravity. We will argue
that we may get over this problem by turning on femrionic fields, which in
the case we are discussing, eleven-dimenional supergravity, amounts to
allow for non-vanishing gravitino field.

The plan of the work is as follows: In section \ref{11dimsugra} we
briefly review simple supergravity in eleven dimensions and in section
\ref{compactification} we describe the general procedure for extracting
solutions. In section \ref{Frub} we explicitly present the class of
solutions with both bosonic and fermionic degrees of freedom and in
   \ref{Pureferm} that with pure fermionic fields. In section
\ref{triviality} we examine the triviality of the obtained solutions, and
finally in section \ref{Conclusions} we conclude and we discuss our
results.

\section{Simple supergravity in eleven dimensions}
\label{11dimsugra}

Let us briefly recall ${\cal{N}}=1$  simple supergravity in 11 dimensions
\cite{Cremmer:1978km,Duff:1986hr,Castellani:1983yg}. The field content of
the theory consists of the vierbein $e{^{A}}_{M}$, a Majorana anticommuting
spin-$\frac{3}{2}$ field $\psi_{M}$ and a completely antisymmetric 3-form gauge
field $A_{KLM}$.
The field equations for the eleven dimensional supergravity read
\cite{Cremmer:1978km}:
\bea
\label{gravitinoeom}
\Gamma^{RST}\hat{D}_{S}(\hat{\omega})\psi_{T}&=&0,\\
\label{Feom}
D_{T}(\hat{\omega})\hat{F}^{TURS}&=&(24)^{-2}\epsilon^{MNPQVWXYUSR}\hat{F}_
{MNPQ}\hat{F}_{VWXY},\\
\label{Gravityeom}
R_{TS}(\hat{\omega})-\frac{1}{2}g_{TS}R(\hat{\omega})&=&\frac{1}{24}\left[
g_ { TS }
\hat{F}_{MNPQ}\hat{F}^{MNPQ}-8\hat{F}_{MNPT}\hat{F}^{MNP}{_{S}}\right],
\eea
where the super-covariant $\hat{D}_{S}$ and covariant derivatives $D_{S}$
are 
\bea
\hat{D}_{S}(\hat{\omega})\psi_{T}&=&D_{S}(\hat{\omega})\psi_{T}+T_{S}{^{
MNPQ}}\hat{F}_{MNPQ}\psi_{T},\\
D_{S}(\hat{\omega})\psi_{T}&=&\partial_{S}\psi_{T}+\frac{1}{4}\hat{\omega}_
{SAB}\Gamma^{AB}\psi_{T}.
\eea
The notation 
\bea
T^{SMNPQ}&=&(12)^{-2}\left(\Gamma^{SMNPQ}-8\Gamma^{[MNP}\eta^{Q]S}\right),
\eea
has been used whereas the super-covariant spin-connection and the super-covariant field strength are
\bea
\label{superspinconn}
&&\hat{\omega}_{MRS}=\omega_{MRS}(e^A_{\ M})+\frac{1}{2}i\left(2
\bar{\psi}_{M}\Gamma_{[S}\psi_{R]}+\bar{\psi}_{S}\Gamma_{M}\psi_{R}\right),
\\
&&\hat{F}_{MNPQ}=F_{MNPQ}-3\bar{\psi}_{[M}\Gamma_{NP}\psi_{Q]},
\eea
respectively, with
\bea
F_{MNPQ}&=&4\partial_{[M}A_{NPQ]}.
\eea
It should be stressed that relation (\ref{superspinconn}) leads to the
non-vanishing effective torsion tensor given by
\bea
{T^A}_{MN}=-i\bar{\psi}_M\Gamma^A\psi_N\, .
\label{torsion}
\eea

In these expressions, early alphabet capital letters $(A,B,...=0,...,10)$
are tangent space indices,
while middle and late alphabet capital letters $(K,L....=0,...,10)$ are 
world indices. Moreover,  the  eleven dimensional tangent space
metric $\eta$ is mostly minus, i.e., $\eta_{AB}={\rm diag}(1,-1...-1)$. Finally, eleven-dimensional $\Gamma$-matrices 
are in the
Majorana representation and they form a purely imaginary representation of
the Clifford algebra in 11 dimensions. More details are given in the Appendix.

The supergravity equations (\ref{gravitinoeom})-(\ref{Gravityeom}) are
invariant under the local
supersymmetry transformations
\bea
\delta e{^{A}}_{M}&=&-i\bar{\epsilon}\Gamma^{A}\psi_{M},\\
\delta
\psi_{M}&=&D_{M}(\hat{\omega})\epsilon-\frac{i}{144}\left(\Gamma{^{NPQR}}_{
M }
+8\Gamma^{PQR}\delta^{N}_{M}\right)\hat{F}_{NPQR}\epsilon,\\
\delta A_{MNP}&=&\frac{3}{2}\bar{\epsilon}\Gamma_{[MN}\psi_{P]},
\eea
where $\epsilon$ is the fermionic supersymmetry transformation parameter.

\section{Compactification}
\label{compactification}

The $N=1$ supergravity in eleven dimensions \cite{Cremmer:1978km} is a
maximal theory and  has offered a rich foundation for studying
compactification mechanisms. 
The investigation of the latter has been limited to find solutions to the classical equations 
of motion where all fermionic fields (here gravitino) vanish. 
 Especially, Freund-Rubin type of vacua
\cite{Freund:1980xh} provide a natural framework and pave the way not only
for the spontaneous compactification of eleven-dimensional
supergravity but also for any higher-dimensional theory. However,  this
compactification mechanism where bosonic degree of freedom are excited, suffers from a huge cosmological constant, that
cannot be eliminated or reduced \cite{Castellani:1983yg}. Thus, we are led
to explore the possibility of compactifications   where in addition
fermionic fields are allowed to be non-vanishing and lead to 4D flat
Minksowski vacum. For the case we are discussing, eleven-dimensional
supergravity, this means that we will excite  the gravitino field. However,
the gravitino is an anticommuting Grassmann field and its classical limit
vanishes, and therefore there cannot be a classical gravitino field.
Nevertheless, it can form bilinears or quadrilinears and so on which have
a classical interpretation. 
We recall that for  an anticommuting   charged fermion field $\Psi(x)$
satisfying the Dirac equation,
there is a classical electron density field $W(x,p)$ given by the Wigner
transform of the 
bilinear $\bar{\Psi}(x)\Psi(y)$, 
which is measurable and plays a role 
in semiconductor modeling. In a sense, this is equivalent of solving the
equations of motion and taking expectation values 
of the fields
involved. Thus, interpreting the gravitino as  a fermionic quantum field,
its vacuum expectation value  will vanish,
\bea
\langle\psi^A\rangle=0\, , \label{psi22}
\eea
which however does not imply that the
vacuum expectation values of various bilinears should vanish as well.  For
example
(\ref{psi22}) may hold but 
certain of its possible bilinears may be non-zero, 
\cite{Weinberg:1995mt,Peskin:1995ev}, i.e., 
\bea
\langle\psi^A\,\Gamma^{B_1}\cdots\Gamma^{B_n}\,\psi^C\rangle\neq0.
\label{psi22b}
\eea

The non-vanishing of the vacuum expectation values
of some of the gravitini billinears has been considered before
\cite{Duff:1982yi}, \cite{Wu:1984wn} in order to find flat Minkowski vacua.
  Here we construct explicitly solutions that realize
(\ref{psi22}),(\ref{psi22b}) and we describe  compactifications of
the internal space in such a way that the four-dimensional cosmological
constant vanishes.

We are interested in finding solutions of the supergravity equations
(\ref{gravitinoeom})-(\ref{Gravityeom}), 
corresponding to a direct product of a seven-dimensional compact space
with a four-dimensional Minkowski spacetime. The standard ansatz for the
compactification metric should then be
\bea
\langle g_{\mu\nu}\rangle &=&\eta_{\mu\nu},\nonumber\\
\langle g_{mn}\rangle &=&g_{mn},\nonumber\\
\langle g_{m \nu}\rangle &=&\langle g_{\mu n}\rangle =0.
\eea
Eleven-dimensional indices have been split as
\bea
M&=&(\mu,m),\nonumber\\
\mu&=&0,1,2,3,\ ({\text{4D world indices}})\nonumber\\
m&=&1,2,...,7\ ({\text{7D world indices}}).\nonumber
\eea
For the super-covariant 4-form field strength $\hat{F}_{KLMN}$, we would like to have  
\be
\label{Fhvev}
\langle \hat{F}_{KLMN}\rangle =0\, .
\ee
This choice may seem rather ad hoc but it is dictated by the fact
that $\hat{F}_{KLMN}$ transforms covariantly under supersymmetry 
and it is the generalization of the $F_{LKMN}=0$ in the pure bosonic case. 
Definitely the ansatz (\ref{Fhvev}) does not imply that  fermion bilinears
and bosonic four-form  field strength
vanish independently. As we will see there are cases where both fermionic bilinears and bosonic field strength 
is not vanishing and they just  cancel each
other. For the gravitino field  we consider the splitting
\be
\psi^{A}=(\psi^{\alpha},\psi^{a}).
\ee
From the four dimensional point of view $\psi^A$ gives rise to 8 spin $3/2$ gravitini $\psi^\alpha$ and
$8\!\times\!7=56$ spin $1/2$ fermions $\psi^a$. 
The next step is to ensure that the torsion is non vanishing  and  fully antisymmetric only in the
internal space, while it is zero in the external
4D spacetime. In the opossite case, a non-zero torsion in 4D, would require a non-flat metric as torsion would act as source in the 
right-hand side of the 4D Einstein equations (\ref{Gravityeom}). 
In fact the latter can be written as 
\bea
R_{MN}(\hat{\omega})=0\, ,  \label{rmn}
\eea
which is clearly solved by
\bea
\hat{\omega}_{ABC}=0. \label{ho}
\eea
This means that the spin connection is given entirely in terms of the gravitino bilinear
\bea
\omega_{ABC}(e^A_{\ M})=-\frac{1}{2}i\left(2
\bar{\psi}_{A}\Gamma_{[C}\psi_{B]}+\bar{\psi}_{C}\Gamma_{A}\psi_{B}\right)
\, .
\eea
As a result, the spin connection may be written simply in terms of the torsion (\ref{torsion}) as
\bea
\omega_{ABC}=-\left(T_{ABC}-T_{ACB}+T_{CAB}\right)
\eea
In other words, the connection is just the Weitzenb\"ock connection
\bea
{\Gamma^{M}}_{KL} = E^K_A \partial_K E^A_L\, .
\eea
This corresponds exactly to teleparallelism of simple supergravity: since super-connection is zero, Riemann curvature tensor
vanishes and one may define parallel 
transport over finite distances and not only in infinitesimal neighborhood.
However, parallelograms do not close 
under parallel transport, a manifestation of torsion.  This is a known fact 
for supersymmetry \cite{Nieu} (and general relativity as well) and has been considered so far as a curiosity rather than a fact of 
fundamental importance.  
Here we use it as a tool generating technique for finding solutions \cite{keh}.

The vanishing of the supercovariant connection (\ref{ho})  simplifies significantly the equations of motion. In
particular, the gravitino equation (\ref{gravitinoeom}) turns now to be
\bea
\Gamma^{ABC}\partial_{B}\psi_{C}=0\, .
\eea
Clearly solutions to the above equation are provided by constant
anticommuting Majorana fermions $\psi_A$, i.e., 
\bea
\psi_A(x)=\psi_A.
\eea
 
The equation for the supercovariant field strength (\ref{Feom}) 
is also  satisfied due to (\ref{Fhvev}). Finally the gravitational
equation (\ref{Gravityeom}) reduces to (\ref{rmn}), which is split into
\bea
&&R_{\mu\nu}(\hat{\omega})=0\, ,\label{R1}\\
&&R_{mn}(\hat{\omega})=0\, , \label{R2}
\eea
and  which  are always satisfied since $\hat{\omega}=0$. However, since we want 4D Minkowski vacuum, (\ref{R1}) should be solved by 
\bea
g_{\mu\nu}=\eta_{\alpha\beta}e^\alpha_{\mu}e^\beta_{\nu}={\rm diag}(1,-1,1,-1)\, , 
~~~\omega_{\mu \alpha\beta}(e^\alpha_{\mu})=0\, ,
~~~~{T^\alpha}_{\mu\nu}=0\, ,
\eea
where $e^\alpha_{\ \mu}$ is the four-dimensional vierbein.
In other words, we are looking for flat Minkowski metric with vanishing connection and torsion. 

What remains to be solved is actually (\ref{R2}). 
In the following we will separately explore two classes of solutions 
 in the context of the above formulation. In particular, the
first class contains both bosonic ($F_{KLMN}$) and fermionic ($\psi_M$) degrees of freedom,
while the second class contains only fermionic  ($\psi_M$) fields.

A final comment concerns supersymmetry.  The supersymmetry transformations in the present framework take the form
\bea
\label{susygravitation}
\delta e{^{A}}_{M}&=&-i\bar{\epsilon}\Gamma^{A}\psi_{M},\\
\label{susygravitino}
\delta \psi_{A}&=&\partial_{A}\epsilon,\\
\label{Fsusy}
\delta A_{ABC}&=&\frac{3}{2}\bar{\epsilon}\Gamma_{[AB}\psi_{C]}.
\eea
Supersymmetry is preserved whenever supersymmetry fermionic charges
annihilate the vacuum. This is equivalent to consider 
the supersymmetry parameter $\epsilon$ as a commuting spinor. Then,
contrary to the bosonic case where all bosonic variations vanish 
(due to vanishing of fermionic field), here all fermionic shifts vanish and
one should consider the bosonic ones. 
The latter do not vanish in general due to non-zero  bilinears.

\section{Freund-Rubin compactification with zero cosmological constant}
\label{Frub}

In this section we will consider a Freund-Rubin ansatz for the three-form
field  in the external space \cite{Freund:1980xh}.  In particular, the four-form  field strength
takes the form
\bea
\label{F4}
&&F_{\alpha \beta \gamma \delta}=6 m_0\,\epsilon_{\alpha \beta \gamma
\delta},\, , ~~~~\alpha,\beta,...=0,1,2,3\, , \\
&&F_{KLMN}=0\, ~~~~\mbox{otherwise},
\eea
where $m_0$ is a constant.
 With a vanishing gravitino field, the field equations for the three-form field and gravity are reduced to the solution of
\bea
&&R_{\mu\nu}=-12 m_0^2\,g_{\mu\nu}\\
&&R_{mn}=6 m_0^2 \,g_{mn}.
\eea
For a maximally symmetric background we find the celebrated $AdS_4\times S^7$ vacuum. 

We will now turn on a gravitino field and look for solutions of the form $M^4\times B^7$ where $M^4$ is a 4D Minkowski spacetime and 
$B^7$ is a 7D compact manifold. Below we will present 
three different solutions for the gravitino field, that solves the supergravity fields equations and 
provide a vacuum of this form.

\subsection{Solution $M^4 \times S^7$}
\label{m4s7}

To proceed, let us consider the following form of the gravitino $\psi_M$
\bea
\label{psialpha1}
\psi^{\alpha}&=&(H^{\alpha}\otimes\theta),\\
\label{psia1}
\psi^{a}&=&(J\otimes\theta^{a}),
\eea
with
\bea
J&=&(a,a,c,c)^{T},\nonumber\\
H^{\alpha}&=&\gamma^{5}\gamma^{\alpha}J,\nonumber\\
\theta&=&(1,0,0,0,0,0,0,0)^{T},\nonumber\\
\theta^{a}&=&\tau^{a}\theta.\nonumber
\eea
The parameters $a,c$ are constant real Grassmann variables, which guarantee that 
the gravitino satisfies the Majorana condition
\cite{VanNieuwenhuizen:1985be}, 
\be
\psi_M^{\dagger}\Gamma^{0}=\psi_M^{T}C\, .
\ee 
 Moreover, it can easily be verified that
\be
\bar{\psi}_{[\alpha} \Gamma_{\beta \gamma}
\psi_{\delta]}=(4iac)\epsilon_{\alpha \beta \gamma \delta},
\ee
 so that the condition $\hat{F}_{KLMN}=0$ gives 
\be
\label{fieldstreansatz}
F_{\alpha \beta \gamma \delta}=(12iac)\epsilon_{\alpha \beta \gamma
\delta}\, .
\ee
Thus, by comparing it with (\ref{F4}), we find that 
\be
m_0=6i\, a c.
\ee
Then also  all conditions at the beginning of section (\ref{compactification}) hold. In particular,  
by using the
relations
\bea
\theta^{T}\tau_{a}\theta&=&0,\nonumber\\
\theta^{T}\tau_{a}\tau_{b}\theta&=&-\delta_{ab},\nonumber\\
\theta^{T}\tau_{a}\tau_{b}\tau_{c}\theta&=&-a_{abc},
\nonumber
\eea
where $a_{abc}$ are the octonionic stracture constants
\be
a_{abc}=-\theta^{T}\tau_{abc}\theta\, ,
\ee
we find that the torsion is 
\bea
\label{Sol1}
&&T_{abc}=\frac{i}{2}\bar{\psi}_{a}\Gamma_{b}\psi_{c}=-\frac{m_0}{3}a_ { abc } , \, , ~~~~a,b,c=1,...,7\\
&&
T_{ABC}=0 \, , ~~~~~~~~\mbox{otherwise}
\eea
Therefore, the torsion is  fully antisymmetric  in the internal space and
it vanishes
anywhere else. 
It leads also (\ref{Sol1}) to the spin connection 
\bea
&& \omega_{abc}(e^a_{\ m})=-\frac{m_0}{3}a_{abc},\label{oo}\\ 
&& \omega_{ABC}=0\, , ~~~~~\mbox{otherwise}\, . 
\eea
In particular, the spin connection vanishes for the external spacetime
\bea
\label{externalspinconn}
\omega_{\alpha\beta\gamma}(e^\alpha_{\ \mu})=0\, .
\eea
Thus,   since the spin connection
(\ref{externalspinconn}) vanishes, the external space  can be chosen to be a four
dimensional Minkowski spacetime ($M^4$)
\bea
g_{\mu\nu}&=&\eta_{\mu\nu},\\
R_{\mu \nu}(e^\alpha_{\ \mu})&=&0.
\eea

Concerning the internal space,  we recall   the
well known result that when the spin connection has the form (\ref{oo}) a solution to eq.(\ref{R2}) 
is provided  by the parallelized  seven-sphere ($S^7$) and thus
\cite{Duff:1982yi,Englert:1982vs} 
\be
R_{mn}(e^a_{\ m})=0.
\ee
In summary, the 11-dimensional spacetime splits as  $M^4 \times S^7$.

\subsection{Solution $M^4 \times S^3 \times T^{4}$}
\label{m4s3t4}

We may continue looking for other
compactifying  spaces with an ansatz similar to
(\ref{psialpha1}),(\ref{psia1}), where this time  $\theta^{a}$ is of the form
\bea
\theta^{a}=(\tau^{1}\theta,\tau^{2}\theta,\tau^{3}\theta,0,0,0,0).
\eea
Then the torsion turns out to be 
\bea
&&T_{\bar{a}\bar{b}\bar{c}}=(-2iac)\epsilon_{\bar{a}\bar{b}\bar{c}}\, , ~~~~~  \bar{a},\bar{b},\bar{c}=1,2,3\\
&&T_{ABC}=0\, , ~~~~~~~\mbox{otherwise} .
\eea
This is just the
parallelizing torsion for the three-sphere ($S^3$). We also have 
\bea
\label{fieldstreansatz1}
F_{\alpha \beta \gamma \delta}=(12iac)\epsilon_{\alpha \beta \gamma
\delta}\, ,
\eea
in order to have $\hat{F}_{KLMN}=0$. Einstein equations are then simply
\bea
&&R_{\mu\nu}=0\, , \label{R11}\\
&&R_{\bar{m}\bar{n}}=0\, , \label{R22}\\
&&R_{mn}=0\, , \label{R33}
\eea
where $m,n=1,...,4$ and $\bar{m},\bar{n}=1,2,3$.
The  solution to (\ref{R11}) is just 4D Minkowski space $M^4$, the solution to (\ref{R22}) is provided by the paralellized $S^3$
and the maximally symmetric solution to (\ref{R33}) is a flat torus $T^4$.

\subsection{Solution $M^{4} \times T^{7}$}
\label{m4t7}
There is another class of solutions where the non-vansihing components of the gravitino is along four dimensions only, i.e., 
\bea
\psi^{\alpha}&=&(H^{\alpha}\otimes\theta),\\
\psi^{a}&=&0.
\eea
A gravitino field of this sort leads to 
\be
\omega_{ABC}(e^A_{\ M})=0\, , 
\ee
 and the
torsion is always zero. This kind of solution
implies a Minkowski external spacetime ($M^4$) and a flat 7D torus $T^7$-geometry in the maximally symmetric case. In
summary the 11-dimensional spacetime is of the form    $M^{4} \times T^{7}$.

\subsection{Supersymmetry}

Let us now investigate the  supersymmetry transformations of the above
three solutions. The first two solutions, namely of subsections
\ref{m4s7} and \ref{m4s3t4}, have completely broken supersymmetry, that is
there is not any spinorial parameter that makes all field-shifts to
vanish. However, for the solution if subsection \ref{m4t7}
 there is one such spinorial parameter $\epsilon$ that makes all
field-shifts to vanish, namely
\be
\epsilon=T\otimes\theta \, , \label{eps}
\ee
where
\bea
T=(-a,a,c,-c)^{T}.
\eea
It is then straightforward to verify that 
\bea
\delta e{^{A}}_{M}&=&0,\\
\delta \psi_{M}&=&0,\\
\delta A_{MNP}&=&0 \ \ ({\text{up  to  a  gauge  transformation}}).
\eea
for supersymmetry parameters $\epsilon$ given in (\ref{eps}). However, as there are non-vanishing fermionic fields, Lorentz
symmetry is broken.

\section{Pure fermionic solutions with zero cosmological constant}
\label{Pureferm}

We will consider here a class of solutions where no antisymmetric three-form field is turned on. As a result, the 
only non vanishing fields are the graviton and the gravitino. 
 We
will present two different solutions for the gravitino field, that will
lead to Minkowski vacuum and thus  to a vanishing  cosmological constant.

\subsection{$M^{4} \times S^{7}$} 
\label{m4s7ferm}

Assume a gravitino of the form 
\bea
\psi^{\alpha}&=&(0,0,0,H^{3}\otimes\theta),\\
\psi^{a}&=&(J\otimes\theta^{a}),
\eea
where
\bea
J&=&(a,a,c,c)^{T},\nonumber\\
H^{3}&=&\gamma^{5}\gamma^{3}J,\nonumber\\
\theta&=&(1,0,0,0,0,0,0,0)^{T},\nonumber\\
\theta^{a}&=&\tau^{a}\theta.\label{fermionicconvent}
\eea
The constants $a$ and $c$ are again anticommuting Grassmann variables and
the Majorana condition is still valid. However, contrary to  the previous class of 
solutions in (\ref{fieldstreansatz}), the condition $\hat{F}_{KLMN}=0$
leads to   
\bea
F_{ABCD}=0,
\eea
since now 
\be
\bar{\psi}_{[M}\Gamma_{NP}\psi_{Q]}=0.
\ee

Note that the vanishing of the field strength is realized in the whole
spacetime, including the external one, that is
$F_{\alpha\beta\gamma\delta}=0$. Moreover, for the torsion we find 
\bea
\label{Sol111}
&&T_{abc}\equiv\frac{i}{2}\bar{\psi}_{a}\Gamma_{b}\psi_{c}  \, , ~~~~a,b,c=1,...,7\\
&&
T_{ABC}=0 \, , ~~~~~~~~\mbox{otherwise},
\eea
so that the spin connection turns out to be  
\bea
&& \omega_{abc}(e^a_{\ m})=-(2iac)a_{abc},\label{oo11}\\ 
&& \omega_{ABC}=0\, , ~~~~~\mbox{otherwise}\, . 
\eea
It is obvious then that the internal space is compactified
to a parallelized seven sphere, and the external space turns out to be a
four dimensional Minkowski space-time. That is the 11D
spacetime splits as $M^{4} \times S^{7}$.

\subsection{$M^{8} \times S^{3}$}
\label{m8s3ferm}

The second solution which we can get with only fermionic condensates
contributing to the ground state, is given by 
\bea
\psi^{\alpha}&=&(0,0,0,H^{3}\otimes\theta),\\
\psi^{a}&=&(J\otimes\tau^{1}\theta,J\otimes\tau^{2}\theta,J\otimes\tau^{3}
\theta,0,0,0,0).
\eea
The conventions are also given by (\ref{fermionicconvent}), but now we have
a parallelized three sphere $S^3$ and an 8D Minkowski spacetime $M^8$. That is
the 11D spacetime compactifies to $M^{8} \times S^{3}$. Of course $M^8$ can further be compactified on a 4D torus $T^4$ 
to $M^4\times T^4$. This corresponds to a $M^{4} \times T^{4} \times S^{3}$ compactification of the 11D theory.

\subsection{Supersymmetry}

Let us now consider the supersymmetry transformations of the solutions of
subsections \ref{m4s7ferm} and \ref{m8s3ferm}.
It is possible to find the existence of three supersymmetries if we use the
following explicit form for the supersymmetry parameter:
\be
\epsilon_{(i)}=T_{(i)}\otimes\theta,
\ee
where
\bea
&&T_{(1)}=(c,c,0,0)^{T},\nonumber\\
&&T_{(2)}=(0,0,a,a)^{T},\nonumber\\
&&T_{(3)}=(a,a,-c,-c)^{T}.
\eea
Then for both solution subclasses:
\bea
\delta e{^{A}}_{M}&=&0,\\
\delta \psi_{M}&=&0,\\
\delta A_{MNP}&=&0\ ({\text{up to a gauge transformation}}).
\eea

\section{Triviallity of the Fermionic Vacua}
\label{triviality}

One question concerns the triviallity of the vacuum solutions found above. Triviality here has the meaning of relating 
these vacua with non-zero fermionic fields to pure bosonic backgrounds with vanishing fermionic fields by means of a supersymmetry 
transformation. For example, for the gravitino profile
(\ref{psialpha1}),(\ref{psia1}),  triviality means that there is 
Majorana spinor $\epsilon$ such that 
\be
\psi_\alpha=D_\alpha(\hat{\omega})\epsilon\, ,
~~~~~\psi_a=D_a(\hat{\omega})\epsilon\label{tr}.
\ee
If (\ref{tr}) is valid, then, $\psi_\alpha,\psi_a$ can be shifted to zero ($\psi_\alpha=0,\psi_a=0$) by means of a 
supersymmetry transfrormation, resulting in a pure bosonic 
background by a corresponding shift of the veilbein and the three-form field. 

Let us suppose that the solution in section \ref{m4s7} is trivial and indeed $\psi_\alpha,\psi_a$ can be written as in (\ref{tr}). This means 
that there is a non-zero $\epsilon$ which we write as 
\be
\epsilon=J\otimes \lambda
\ee
and thus, $\lambda$ should satisfy for example
\be
D_a(\hat{\omega})\lambda=\tau_a \theta. \label{thet}
\ee
Since $\hat{\omega}=0$, we find that $\lambda$ should satisfy the condition
\be
\tau^m\partial_m\lambda=7 \theta.
\ee
Acting with $\tau^m\partial_m$ on both sides of the above equation, we get 
\be
\Box \lambda_i=0 \label{ll},
\ee
where $\lambda_i$ are c-functions in the expansion of  $\lambda$ in terms of a Grassmann base $(a_i)$, $\lambda=\sum_i\lambda_i a_i$.
Multiplying (\ref{ll}) with $\lambda^*$ and integrating over the whole $S^7$, we find that $\lambda_i={\rm const.}$ and 
similarly $\epsilon$ is  a constant spinor.  Then clearly
there is no solution with constant $\epsilon$ to (\ref{thet}) and thus the solution for $(\psi_\alpha,\psi_a)$ is not trivial.

Proceeding similarly, one may prove that none of the solutions presented is
trivial except the $M^4\times T^7$ of section 
(\ref{m4t7}). The latter has no paralelizing torsion at all and it can be seen that the gravitino is pure gauge 
(connected to $\psi_M=0$ by a susy transformation).

Below we review the solutions found with the help of the following
table

\begin{center}
  \begin{tabular}{ |c | c | c |  c|}
    \hline
    fermionic v.e.v.s & bosonic v.e.v.s & geometry  &triviallity\\
    \hline
\hline
    Majorana & yes & $M^4 \times S^7$ & no \\ \hline
    Majorana & yes & $M^4 \times S^3 \times T^{4}$ &no\\
    \hline
    Majorana & yes & $M^4 \times T^{7}$  & yes
 \\ \hline
    Majorana & no & $M^4 \times S^7$  & no\\
    \hline
    Majorana & no & $M^8 \times S^3$  & no\\
    \hline
  \end{tabular}
\end{center}

\section{Conclusions}
\label{Conclusions}

As has been stressed above, all known compactifications of higher dimensional theories (supersymmetric or not) inherit a cosmological constant to 
4D spacetime. Although this is not the case for a supersymmetric compactification, like  a $CY$ or any other appropriate 
internal space, a cosmological constant emerges as soon as supersymmetry is broken. In the latter case, the value of the 
cosmological constant is determined by the gravitino mass, which is nevertheless non-zero (or many orders of magnitude away 
of the reported value of the cosmological constant). On the other hand, 
for non-supersymmetric compactifications the cosmological constant value is determined by the size of the internal space. 
In any case, standard compactifications produce a huge non-zero cosmological constant. The word ``standard'', refers here to 
the well established procedure of putting all fermionic fields to zero and allowing for non-zero bosonic fields to determine 
the vacuum of the theory. There is nothing wrong with this as long as a mechanism to reduce the cosmological constant 
is found. But one would like to explore the possibility of other compactifications which do  not inherit a cosmological constant 
in first  place. 

In the search for such compactifications, we are led to the conclusion that one may allow for non-vanishing fermionic fields,
which in the case of 11D supergravity translates into a non-vanishing
gravitino field. Of course, in this case, one faces the fact that
there is no classical limit of a fermionic field, strictly speaking, as the
latter is an anticommuting object. However, recalling
that although a fermion field cannot be classical,  fermion bilinears,
quadrilinears and so on may have classical limits 
and therefore may be observed. Hence, the way we are to intepret our
findings is that the vacuum of the theory is determined by 
the expectation values of the fields involved and their local products. Thus, although we find a non-vanishing profile 
for the 11D gravitini, the classical vacuum is determined by its expectation value of itself and its local products like 
$\langle\bar{\psi}_M \psi^M\rangle,~\langle\bar{\psi}_M
\Gamma^{M}\psi_N\rangle$ etc. In this way, the backgrounds found do not
produce a cosmological 
constant in 4D, and thus a Minkowski vacuum is possible, although
supersymmetry is clearly broken. As usual nothing comes for free. The cost
paid 
in our case is the appearance of $\psi$-torsion instead of a cosmological
constant (or curvature). 
However, the torsion appears in the internal space and it is responsible
for the parallelism of the internal space and the flatness of
the external, allowing this way Minkowski spacetime as 4D vacuum.      
Indeed in all backgrounds found here,  it is in fact the explicit
solution to the  Majorana condition on the gravitino that
provides the parallelizing torsion in the internal space for each
case, apart from the case $M^4 \times T^{7}$, where no parallelization
exists. The latter is trivial in any case as it is connected by a supersymmetry transformation to a pure bosonic background. 
In  fact appropriate supersymmetry transformation turns $\psi_M=0$ while keeping the $M^4\times T^7$ geometry without vanishing 
three-form field.  

Finally, we expect that more solutions may exist in the spirit of
section \ref{compactification} for eleven dimensional supergravity as well as for lower-dimensional ones, which should be studied.

\begin{acknowledgments}
 This work is supported by the NTUA PEVE-2009 programme for basic research. 
\end{acknowledgments}

\vskip .3in
\noindent
{\bf \Large Appendix}
 
\appendix

\section{Clifford Algebra}


The usual, pure imaginary, Majorana representation of the Clifford algebra
in 11D Kaluza-Klein supergravity is
\cite{Cremmer:1978km,VanNieuwenhuizen:1985be}:
\be
\{\Gamma^{A},\Gamma^{B}\}=2\eta^{AB}\one_{32},
\ee
where $\one_{n}$ is the $n\times n$ unit matrix.  $\Gamma^{M_{1}M_{2}...M_{N}}$ as usual represents
the full antisymmetrized product of $N$ $\Gamma$ matrices.
One convenient representation of the $\Gamma$-matrix algebra is the
following:
\bea
\Gamma^{A}&=&(\Gamma^{\alpha},\Gamma^{a}),\nonumber\\
\Gamma^{\alpha}&=&\gamma^{\alpha}\otimes\one_{8},\nonumber\\
\Gamma^{a}&=&\gamma^{5}\otimes\tau^{a},\nonumber\\
C_{11}=C_{4} \otimes
C_{7}&=&\gamma^{0}\otimes\one_{8}=\Gamma^{0}\nonumber,
\eea
where
\bea
A&=&(\alpha,a),\nonumber\\
\alpha&=&0,1,2,3,\ ({\text{tangent spacetime}})\nonumber\\
a&=&1,2,...,7\ ({\text{tangent internal space}})\nonumber
\eea
and
\bea
(\tau^{a})_{bc}&=&a_{abc},\nonumber\\
(\tau^{a})_{0b}&=&-(\tau^{a})_{b0}=\delta_{ab},\nonumber\\
(\tau^{a})_{00}&=&0.\nonumber
\eea
In the above expressions, the $\tau$ matrices form a real representation of
the seven
dimensional Clifford algebra:
\be
\{\tau^{a},\tau^{b}\}=-2\delta^{ab}\one_{8}.
\ee
Additionally,  $a_{abc}$ are the octonionic algebra
full antisymmetric structure constants \cite{VanNieuwenhuizen:1985be},
which vanish apart from the following entries:
\be
a_{147}=a_{123}=-a_{156}=a_{257}=a_{246}=a_{367}=-a_{345}=1.
\ee

The four dimensional $\gamma$ matrices are considered in the Majorana
representation and form a pure imaginary representation of the clifford
algebra in a minkowski space:
\be
\{\gamma^{\alpha},\gamma^{\beta}\}=2\eta^{\alpha \beta}\one_{4},
\ee
where $\eta^{\alpha\beta}$ has the form diag[1,-1,-1,-1]. They explicitly
have the form:
\bea
\gamma^{0}&=&\left(\begin{tabular}{llrr}
    0 & 0 & 0 & -i \\
    0 & 0 & i & 0 \\
    0 & -i & 0 & 0 \\
    i & 0 & 0 & 0 \\
\end{tabular}\right),\\
\gamma^{1}&=&\left(\begin{tabular}{llrr}
    i & 0 & 0 & 0 \\
    0 & -i & 0 & 0 \\
    0 & 0 & i & 0 \\
    0 & 0 & 0 & -i \\
\end{tabular}\right),\\
\gamma^{2}&=&\left(
  \begin{tabular}{llrr}
    0 & 0 & 0 & i \\
    0 & 0 & -i & 0 \\ 
    0 & -i & 0 & 0 \\
    i & 0 & 0 & 0 \\ 
\end{tabular}\right),\\
\gamma^{3}&=&\left(
  \begin{tabular}{llrr}
    0 & -i & 0 & 0 \\
    -i & 0 & 0 & 0 \\ 
    0 & 0 & 0 & -i \\
    0 & 0 & -i & 0 \\ 
\end{tabular}\right),\\
\gamma^{5}&=&\left(
  \begin{tabular}{llrr}
    0 & -i & 0 & 0 \\
    i & 0 & 0 & 0 \\ 
    0 & 0 & 0 & i \\
    0 & 0 & -i & 0 \\ 
\end{tabular}\right).
\eea

\providecommand{\href}[2]{#2}

\begingroup

\raggedright

\endgroup

\end{document}